\documentclass[aoas,MSNbibl,nameyear,dvips]{arximspdf}
\usepackage{dcolumn}
\usepackage{graphicx}

\doi{10.1214/13-AOAS688} %kopijuoti is PTS
\volume{8}
\issue{1}
\pubyear{2014}
\firstpage{331}
\lastpage{351}

\makeatletter
\newcolumntype{d}[1]{D{.}{.}{#1}}
\makeatother

\begin{document}
\begin{frontmatter}

\title{A semi-parametric Bayesian model of inter- and intra-examiner
agreement for periodontal probing~depth}
\runtitle{Examiner agreement for periodontal probing depth}

\begin{aug}
\author[A]{\fnms{E.~G.} \snm{Hill}\corref{}\thanksref{t1}\ead[label=e1]{hille@musc.edu}}
\and
\author[B]{\fnms{E.~H.} \snm{Slate}\thanksref{t2}\ead[label=e2]{slate@stat.fsu.edu}}
\runauthor{E.~G. Hill and E.~H. Slate}
\affiliation{Medical University of South Carolina and Florida State University}
\address[A]{Department of Public Health Sciences\\
Medical University of South Carolina\\
Hollings Cancer Center\\
86 Jonathan Lucas Street\\
Suite 118 MSC 955\\
Charleston, South Carolina 29425-9550\\
USA\\
\printead{e1}} %adresu isvedimo komanda gale!
\address[B]{Department of Statistics\\
Florida State University\\
117 North Woodward Avenue\\
P.O. Box 3064330\\
Tallahassee, Florida 32306-4330\\
USA\\
\printead{e2}}
\end{aug}
\thankstext{t1}{Supported in part by NIH Grants K25DE016863,
P20RR017696, U24DE016508 and P30CA138313.}
\thankstext{t2}{Supported in part by NSF Grant DMS0604666, and by NIH
Grants P20RR017696, R01DE16353, R03DE020114, R03CA137805 and U24DE016508.}

% HISTORY:
\received{\smonth{4} \syear{2011}}
\revised{\smonth{9} \syear{2013}}

% ABSTRACT
%
\begin{abstract}
Periodontal probing depth is a measure of periodontitis severity.
We develop a Bayesian hierarchical model linking true pocket depth
to both observed and recorded values of periodontal probing depth,
while permitting correlation among measures obtained from the same mouth
and between \mbox{duplicate} examiners' measures obtained at the same
periodontal site.
Periodontal site-specific examiner effects are modeled as arising from
a Dirichlet
process mixture, facilitating identification of classes of sites that
are measured with similar bias.
Using simulated data, we demonstrate the model's ability to recover
examiner site-specific bias and variance heterogeneity and
to provide cluster-adjusted point and interval agreement estimates.
%We further demonstrate our model's ability to correct for biased
%estimation of chance-corrected agreement via %simultaneous modeling of
%intra- and inter-examiner variation.
We conclude with an analysis of data from a probing depth calibration
training exercise.
\end{abstract}

% KEYWORDS
\begin{keyword}
\kwd{Agreement}
\kwd{cluster-correlated data}
\kwd{clustering}
\kwd{Dirichlet process mixture model}
\kwd{measurement error}
\kwd{periodontal disease}
\kwd{weighted kappa}
\end{keyword}

\end{frontmatter}

%s1 #&#
\section{Introduction}\label{sec1}

Periodontitis is a chronic infectious disease characterized
by gingival bleeding and attachment loss, an increase in
pocket depth (the distance from the gingival crest to the base of the
periodontal pocket), and bone loss.
%Twenty-six percent of U.S. adults age
%20 years and older suffer from destructive periodontitis (defined as
%having attachment loss of at least 4mm at one or more sites), with the
%greatest disease burden
%experienced by the poor, those who are less educated, and non-Hispanic
%blacks (\cite*{DRC2002}).
Periodontitis is diagnosed using measures of clinical attachment loss
and probing depth,
and the present analysis concerns examiner agreement with respect to
the latter.

%s1.1 #&#
\subsection{Data description}\label{sec1.1}
The motivating data were obtained from a
calibration exercise for dental hygienists in the clinical core of the
South Carolina
Center of Biomedical Research Excellence for Oral Health at the Medical
University of South Carolina.
These data are from a pilot calibration
study used to obtain preliminary measures of
agreement and corresponding uncertainty. Results were used subsequently
to design a formal examiner calibration study described elsewhere
[\citet{Hill2006}].

%Probing depth, the clinical measure of pocket depth,
%is measured using a manual probe.
%at up to six periodontal sites per tooth:
%the distobuccal, midbuccal, mesiobuccal, distolingual, midlingual, and
%mesiolingual.
%Buccal sites are those nearest
%the cheek or lips, lingual sites are those nearest the
%tongue, and distal and mesial sites are those farthest from
%and closest to the midline of the dental arch, respectively.
Periodontitis is a periodontal site-specific disease, meaning that
one site may be severely affected, while a neighboring site
on the same tooth remains unaffected. For this reason, pocket depth is measured
using a manual probe at six sites on the same tooth:
the distobuccal, midbuccal, mesiobuccal, distolingual, midlingual, and
mesiolingual.
Buccal sites are those nearest
the cheek or lips, lingual sites are those nearest the
tongue, and distal and mesial sites are those farthest from
and closest to the midline of the dental arch, respectively.
To facilitate development of the model presented in Section~\ref{ModelDesc},
we distinguish between the following quantities:
(1) \emph{pocket depth}, the true biological state; (2) \emph
{observed probing depth}, the manually probed measure of pocket depth
observed on a continuum; and (3) \emph{recorded probing depth}, equal
to the greatest integer less than or equal to the observed probing
depth. The collection of recorded probing depths comprise the recorded
data for the purposes of analysis.

Prior to formal examiner calibration, a pilot calibration exercise was
devised to provide initial assessment of examiners' performance and identify
areas in which examiners required additional training.
In this study,
a standard examiner~(S)---an experienced periodontist with extensive training in periodontal
research techniques---provided initial training for each of three dental hygienists (A, B and
C) in basic methodology
for clinical research and correct procedures for performing
standardized periodontal examinations.
The pilot calibration study was designed so that
the standard and all hygienists measured pocket depth at all six
periodontal sites
of all teeth except third molars and teeth restored with implants.
%Pocket depth was measured using a University of North Carolina probe,
%and p
Probing depth was recorded as the
largest whole millimeter less than or equal to the value observed on a
manual probe, with
minimum and maximum allowable probing depth measures of 0~mm and 15~mm,
respectively.

A randomization sequence was used to assign examiner pairs
to all quadrants---upper right, upper left, lower left and lower right---of calibration subjects.
This scheme guaranteed examiners evaluated an equal number of quadrants
from the upper and lower arches, and right and left sides.
%Duplicate
Probing depth measurements were obtained from nine subjects,
and both inter- (AS, BS, CS, AB, AC and BC) and intra-examiner (AA, BB,
CC and SS)
data were collected.
Each site was probed by exactly two examiners since
pocket depth may increase with additional repeat probings [\citet
{Osborn1992}].
Measures of a site by the same examiner contributed to intra-examiner
assessment, while those from
different examiners were used to evaluate inter-examiner agreement.
Thus, a fully dentate calibration subject contributes 336 site-level
measurements
(28 teeth${}\times{}$6 sites per tooth${}\times{}$2 examiner measurements
per site)
from which we obtain examiner-pair calibration indices reflecting
agreement at the level of the periodontal site.

%s1.2 #&#
\subsection{Measuring agreement}\label{sec1.2}

%It is well-accepted within the periodontal research community
%that analyses of site-level periodontal measurements should
%account for within-subject correlation
%(\cite*{Blomqvist1985}, \cite*{Fleiss1988}, \cite*{Hoberman1992},
%What is less intuitive is the potential for within-subject correlation
%to exist
%among site-level binary indices of agreement, that is, correlation
%among subject-level collections
%of zeroes and ones, the latter indicating agreement and former
%disagreement between examiners' measurements.
In a separate analysis of these data, \citet{Hill2006}
demonstrate the need to account for within-subject correlation among
site-level binary indices of agreement.
They report cluster-adjusted point and interval estimates of
percent exact agreement and agreement within 1~mm for probing depth,
%with corresponding design effects ranging from 0.03 to 6.8 for
%different examiner pairs.
with confidence intervals constructed as described by Cochran
[\citet{Cochran1977}, pages 240--270].
%They determine that d]isagreement with the standard examiner is
%significantly associated
%with deeper periodontal pockets, and conclude that the correlation
%among periodontal site-level agreement
%indices is attributable in part to the clustering of deep pockets
%within the same mouth.
%Adjusting the estimated variance of a proportion for the effects of
%clustering is straightforward
%(see \cite*[pp. 240--246]{Cochran1977}).
%Based on the asymptotic normality of the sample proportion, $p$, they
%construct
%95\% confidence intervals using $p \pm1.96 \times\mbox{SE}_{\mbox{
%where $\mbox{SE}_{\mbox{\scriptsize{cluster}}}(p) = [\mbox{Var}_{\mbox{
%$\mbox{Var}_{\mbox{\scriptsize{cluster}}}(p)$ is the variance of the
%sample proportion adjusted for the clustering of sites within mouth.
%%Specifically,
%$\mbox{Var}_{\mbox{\scriptsize{cluster}}}(p) =
%where $k$ is the total number of clusters (here, mouths), $i$ indexes
%the clusters, $i = 1, \ldots, k$,
%$m_i$ is the size of the $i$th cluster (here, number of sites), $p_i$
%is the sample proportion for the $i$th cluster, and %$m_{\mbox{
%240--247]{Cochran1977}.
For this pilot calibration data, the asymptotics of the variance
estimator for the cluster-adjusted sample proportion are compromised
due to two dominant data characteristics:
(1) the number of clusters (subjects) for a~given examiner pair is
small (range${}={}$1 to 5); and (2) for a given examiner pair,
the cluster sizes are large relative to the number of clusters (average
cluster size${}={}$35 periodontal sites).
Of the 20 cluster-adjusted 95\% confidence intervals for percent
agreement (exact or within 1~mm)
of probing depth reported by \citet{Hill2006},
one is truncated at 0\%, nine are truncated at 100\%, two are
nonestimable because the point estimate
is constructed from a single cluster, and one is nonestimable because
the within-cluster estimates of agreement
are equal.
%Even non-parametric variance estimation methods, such as the
%bootstrap, are
%not viable approaches here as a consequence of the
%small number of clusters.
%Finally, w
We note that estimating the uncertainty associated with any agreement measure
(e.g., weighted kappa or intra-cluster correlation)
is complicated by these data limitations.

Other authors have addressed the issue of agreement estimation for
correlated observations.
\citet{Williamson1997} use a latent variable model to assess
examiner agreement in classifying cervical
ectopy where the examiners use two different cervical assessment
methods. Their model includes random effects to capture both the
correlation among ratings from the same subject using different
assessment methods, as well as the correlation
between ratings obtained from the same examiner using different
assessment methods.
In another paper, \citet{Williamson2000} analyze this same data
using a pair of generalized estimating equations (GEEs),
the first modeling the marginal distribution
of the ratings and the second modeling the binary indicator of
agreement between two subject-level ratings.
\citet{Oden1991} tackles the problem of agreement estimation for
correlated binocular ratings.
He derives an expression for the approximate variance of a
pooled-$\kappa$
estimate for paired left- and right-eye ratings under the
assumption that the true left- and right-eye $\kappa$ values are the same.

For our data, we use a Bayesian hierarchical modeling approach and specify
three separate but conditionally related
models for: (1) pocket depth, with subject-specific random effects capturing
the marginal correlation among pocket depths within the same mouth; (2)
observed probing depth
conditional on pocket depth, with marginal
correlation between duplicate observations from the same periodontal
site, and a Dirichlet process prior (DPP)
on the examiner-bias parameters to
accommodate possible latent class structure in examiner effects; and (3)
recorded probing depth conditional on observed probing depth, from
which the data likelihood is constructed.
We simulate data from the posterior predictive distribution to
estimate indices of agreement and to obtain corresponding interval estimates
corrected for the correlation in these data.

The motivation for our approach is twofold.
First, a model-based approach utilizing all the data facilitates
borrowing of strength across examiner pairs
and helps mitigate problems associated with small numbers
of large clusters (here, nine clusters with maximum size of 336).
%However, even when considered collectively, the data are still
%characterized by large cluster sizes (maximum cluster size = 336) and
%a small number of clusters ($n$ = 9),
%a feature that potentially compromises the asymptotics of alternative
%modeling approaches for correlated data such as GEEs (\emph{need a
%reference}).
Second, by placing the DPP on model parameters, factors associated with
examiner bias need not be specified \emph{a priori}. Rather,
our model learns characteristics associated with bias, and allows these
to vary by
site, subject and examiner.
Investigating such effects in more traditional modeling settings would
require specifying in advance all potential main and interaction effects
of interest.
%%%%%%%%%%%%%%%%%%%%%%%%%%
%LOOK AT DISCUSSION TO MAKE SURE THERE IS NO REDUNDANCY
%%%%%%%%%%%%%%%%%%%%%%%%%%
We describe our model in Section~\ref{ModelDesc} and
present results from a simulation study in Section~\ref{simstudy}. We
apply our method to the
periodontal probing depth calibration study data and summarize our
results in Section~\ref{application}.
We conclude in Section~\ref{discussion} with a discussion of the
merits of our approach and
identify areas for further research.

%s2 #&#
\section{Model specification}\label{sec2}
\label{ModelDesc}
We consider recorded probing depths to be measurements that result from
error-prone and biased observations made of unobservable true pocket
depth. We therefore construct our hierarchical model by sequentially
modeling these phenomena, from truth to data.

%s2.1 #&#
\subsection{Pocket depth}\label{sec2.1}
\label{PocketDepth}
Among U.S. adults, probing depth follows a
positively skewed distribution with the majority of values
falling in the 1~mm to 3~mm range; depths greater
than 6~mm occur infrequently
[\citet{Albandar1999}].
Based on this observation and assuming all pockets have positive depth,
%Assuming all pockets have positive depth, and based on the
%distribution's shape,
we model pocket depth
% - the true state measured by probing depth -
using lognormal distributions as described below.

For each of $n$ subjects,
$m_i$ periodontal sites are examined, where $i = 1, \ldots, n$
and $m_i$ is the total number of sites examined across all teeth for
subject $i$.
Let $\theta_{ij}$ be the pocket depth for the $j$th
site of the $i$th subject,
$j = 1, \ldots, m_i$. We model the marginal correlation among
pocket depths from the same subject
using random effects. Specifically, we write
%e1 #&#
\begin{eqnarray}
\label{pocketdepthmodel} \log(\theta_{ij}) &= &\mu+ b_i +
\varepsilon_{ij},
\end{eqnarray}
where
%$\log(\theta_{ij})| ~\mu, b_i,\sigma^2_{\varepsilon}
%&
%&
$b_i|\sigma^2_b
%&
\sim
%&
\mathrm{N}(0,\sigma^2_b)$,
$\varepsilon_{ij}|\sigma^2_{\varepsilon}
%&
\sim
%&
\mathrm{N}(0,\sigma^2_{\varepsilon})$,
and $b_i$ and $\varepsilon_{ij}$ are assumed independent.
This model yields an exchangeable correlation structure in which all
sites in
the same mouth are equally correlated, a simplifying assumption that
has been used previously
in the analysis of periodontal data [\citet{Derouen1991}].
%N$(\mu,\sigma^2)$ indicates a normal distribution with mean $\mu$ and
%variance $\sigma^2$.
%s2.2 #&#
\subsection{Observed probing depth}\label{sec2.2}
\label{probingdepth}
Like pocket depth, we assume observed probing depth is positive and
follows a lognormal distribution.
Let $k = 1, 2$ index the duplicate observed probing depth measures,
$T_{ijk}$, for the $j$th periodontal site of subject~$i$.
Because $T_{ijk}$ is the probing depth observed by any one of the four
examiners, we introduce indicator variables to
denote the examiner associated with observation $T_{ijk}$.
Hygienists' performance relative to the standard is of primary
importance, and
we therefore select the standard as the reference level for the
examiner indicator variables.
%For each measure, let $E$ index the examiner, A, B, C or S, making
%that observation.
Accordingly, let
%we define examiner-site specific indicator variables
$X_{E,ijk}$ equal 1 when $T_{ijk}$ is measured by examiner $E ={}$A, B
or C, and 0 otherwise.
Then $\mathbf{X}_{ijk} = (X_{A,ijk}, X_{B,ijk}, X_{C,ijk})^{\prime}$
consists of all zeros if the examiner is the standard,
and is a vector of zeros and a single one otherwise.
Let $\beta_{E,ij}$ be the effect, relative to the standard examiner,
of examiner $E$ on observed probing depth, so that
$\bolds{\beta}_{ij} = (\beta_{A,ij},\beta_{B,ij},\beta
_{C,ij})^{\prime}$ is the parameter
vector associated with
$\mathbf{X}_{ijk}$.
%$\bX_{ijk} = (X_{A,ijk}, X_{B,ijk}, X_{C,ijk})^{\prime}$.
%Note that because S is the reference level for the examiner indicator
%variables,
%the covariate vector
%$\bX_{ijk}$ consists of all zeros if the examiner is the standard,
%and is a vector of zeros and a single one otherwise.

Let $\mathbf{T}_{ij} = (T_{ij1}, T_{ij2})^{\prime}$
%be duplicate observed probing depth measures for the
%$j$th site of the $i$th subject. We
and model $T_{ijk}$
%, k = 1,2$
as
%e2 #&#
\begin{eqnarray}
\label{observedmodel} \log(T_{ijk}) &= &\log(\theta_{ij}) +
\mathbf{X}_{ijk}^{\prime
}\bolds{\beta}_{ij} +
\gamma_{ijk}.
\end{eqnarray}
%
%where
%$\log(T_{ijk})|\log(\theta_{ij}),\bbeta_{ij},\sigma^2_{E}
%&
%&
%$\gamma_{ijk} | \bX_{ijk}, \sigma^2_{E}
%&
%&
%Note that $T_{ijk}$ requires no examiner subscript since the
%combination of $i$, $j$ and $k$ uniquely determines examiner by way of
%the indicator variables.
%To accommodate variance heterogeneity across examiners,
To accommodate variance heterogeneity across examiners, we model the
error terms
$\gamma_{ijk}|\mathbf{X}_{ijk}, \bolds{\sigma}^2 \sim\mathrm{N}(0,
\tilde{\mathbf{X}}_{ijk}^{\prime} \bolds{\sigma}^2)$,
where $\tilde{\mathbf{X}}_{ijk} = (\mathbf{X}_{ijk}^{\prime}, \prod_E
(1 - X_{E,ijk}))^{\prime}$ and $\bolds{\sigma}^2 = (\sigma_A^2,
\sigma_B^2, \sigma_C^2, \sigma_S^2)^{\prime}$.
Thus,
the $\gamma_{ijk}$ are independent mean zero Gaussian random variables
with variance one of $\sigma^2_{A}$, $\sigma^2_{B}$, $\sigma^2_{C}$
or $\sigma^2_{S}$ according to the
examiner associated with observation $T_{ijk}$.
%Hereafter, we use $\sigma^2_{E}$ with $E = $ A, B, C or S to capture
%this variance.
%Note that here the
%examiner subscript of $\sigma^2_{E}$ is A, B, C or S depending on the
%examiner associated with the $k$th
%duplicate of the $j$th site for the $i$th subject.
We further assume that $\gamma_{ijk}$ and
$b_i$, and $\gamma_{ijk}$ and $\varepsilon_{ij}$ are independent.
Finally, since $\theta_{ij}$ is a random quantity, duplicate
observations ($T_{ij1}$ and $T_{ij2}$)
are marginally correlated.

Throughout, we assume the standard examiner exhibits no bias.
%($\beta_{S,ij} = 0$).
However,
if unbiased measuring behavior for the standard cannot be assumed, then
equation (\ref{pocketdepthmodel})
represents ``truth'' as seen by the standard.
Our reference model, Model~0, has common examiner variances and no
examiner biases.
Specifically,
$\sigma^2_{A} = \sigma^2_{B} =
\sigma^2_{C} = \sigma^2_{S} = \sigma^2$ and
$\bolds{\beta}_{ij} \equiv\mathbf{0}$. We consider three
alternative models for observed probing depth
described as follows.

%We further assume the standard's observations
%are subject only to measurement error,
%while we allow for both additional noise and systematic bias in
%hygienists' observations.
%All models considered employ the following common notation.

%s2.2.1 #&#
\subsubsection{Unequal variances and no biases}\label{model1}
Model~1 assumes $\bolds{\beta}_{ij} \equiv\mathbf{0}$, but imposes
no constraints among the examiner variances. Here examiners may differ
in the variability of their probing depth measures, but all are unbiased.
%$\sigma^2_{A} \neq\sigma^2_{B} \neq
%Here differences in observed probing depth among
%examiners is attributable entirely to differences in the variability
%with which examiners obtain probing depth measures, and all measures
%are centered around the truth.

%s2.2.2 #&#
\subsubsection{Unequal variances and constant bias}
\label{model2}
Model 2 relaxes Model 1 by permitting a common examiner effect so that
$\beta_{E,ij} = \beta_{E}$ for all $i$ and $j$.
%In addition to the heterogeneous variance assumption,
%a slightly more complex model, Model 2,
%assumes fixed examiner effects so that
%$\beta_{E,ij} = \beta_{E}$ for all $i$ and $j$.
Thus, study examiners A, B and C are equally biased for all
periodontal sites, but need not have the same bias.
%We placed noninformative Normal$(0,0.001)$ priors on each of
%$\beta_{A}, \beta_{B}$ and $\beta_{C}$.

%s2.2.3 #&#
\subsubsection{Unequal variances and site-level biases}
\label{DPPdescription}
Model 3 further relaxes assumptions by placing a nonparametric
Dirichlet process prior (DPP) on $\beta_{E,ij}$
that supports different effects for examiners A, B and C associated with
site-level characteristics. Our motivation is to
incorporate flexibility in the examiner bias parameters to facilitate
discovery of any latent class structure
among the sites. Here classes define collections of sites with common
measurement bias, the
identification of which may be useful in designing follow-up training
for examiners or future
calibration exercises.
%Classes may be related to periodontal site-specific characteristics
%(e.g. deep pockets or lingual sites), experimental
%design (e.g. sites measured earliest in the day), or subject-level
%characteristics (e.g. sites in older subjects).
Our approach is similar to that used by \citet{Congdon2007} in
which he placed DPPs on
regression coefficients in a bivariate
analysis of male and female suicides in England
to capture spatial variability in the regression parameters.

Escobar and West [\citeauthor{Escobar1995} (\citeyear{Escobar1995,Escobar1998})]
describe mixture modeling via DPPs.
\citet{Congdon2001}, pages~260--273, summarizes their work (and
others) and provides examples of DPP mixture modeling using WinBUGS
[\citet{Lunn2000}].
To briefly summarize, let $y_{i}$, $i = 1, \ldots, n$, be drawn from
the distribution $f(y_{i} | \phi_{i})$, where the
parameter $\phi_{i}$ is unknown.
The DPP treats the underlying distribution of
$\phi_{i}$ as unknown but centered around
a base distribution, $G_0$, from which candidate values for $\phi_{i}$
are drawn
according to a concentration parameter, $\alpha$.
The $\phi_{i}$ cluster based on similarities among the
$y_{i}$, so that assignment of a given candidate value from $G_0$ to
multiple $\phi_{i}$ may be expected.
In practice, $M \leq n$ candidate values, denoted $\phi_{m}^{\ast}$
with $m = 1, \ldots, M$, are drawn from $G_0$
and $M^{\ast} \leq M$ of these are
allocated to one or more of the $\phi_{i}$.
%The density of $\phi_{i}$ is discrete, and the fineness of the
%discretization increases with $\alpha$.
%Specifically,
The density of $\phi_{i}$ more closely resembles $G_0$ for large
values of $\alpha$,
while small values of $\alpha$ result in a density similar to a finite
mixture model.
A practical approach to implementation of the DPP using a~`stick-breaking prior' is described by \citet{Ishwaran2001} and
is based on a finite version of the constructive definition introduced by
\citet{Sethuraman1994}. Our implementation uses this finite approximation.

In our probing depth application, we
accommodate different latent class structures across examiners A, B and
C, and
assign the DPP to the model regression parameters, $\beta_{E,ij}$, as
%follows:
$\beta_{E,ij}|\Gamma_E
%&
\sim
%&
\Gamma_E$ with
$\Gamma_E
%&
\sim
%&
DP(\alpha_E G_{E,0})$.
%,
We specify $G_{E,0}$ to be a normal distribution with examiner-specific
mean and variance,
%= \mathrm{N}(\mu_E,\xi^2_E)$
and $\alpha_E$ is a precision parameter.
Sites $(i, j)$ and $(i^{\prime}, j^{\prime})$ are identified as
belonging to the same cluster for
examiner $E$ when $\beta_{E,ij} = \beta_{E,i^{\prime}j^{\prime}}$.
\citet{Congdon2001} notes
the number of classes cannot exceed the number of distinct data values.
In our application,
recorded probing depth takes integer values ranging from 0~mm to 8~mm,
with the difference in duplicate
measures ranging from $-$4~mm to 4~mm.
These values lead to at most 9 distinct intervals for
the observed probing depth and, although $T_{ijk}$ may vary within
these intervals, such variation within 1~mm adds little to
understanding examiner performance.
%Therefore, the number of classes capturing biases among examiners
%(via the $\beta_{E,ij}$)
%can not exceed nine.
%Furthermore, the number of classes associated with a specific examiner
%is likely to be small given
%the examiners' uniform training.
Since we place separate DPPs on the distribution
of $\beta_{E,ij}$ for examiners A, B and C, we use an examiner
subscript for the number of candidate values, $M_E$,
drawn from the baseline distribution, $G_{E,0}$.
We used $M_E = 6$ for all
examiners to facilitate sufficient model flexibility to discover latent
class structure.
We specified a range of potential values
for the concentration parameters with $\alpha_E = 0.5, 1, 2, 3, 4, 5,
6, 7, 8, 9, 10$ and $20$ and
conducted a sensitivity analysis (summarized in Section~\ref{simulationmodeldescription}) to facilitate selection.
Based on comparisons of posterior class inference across the range of
$\alpha_E$ values,
we selected $\alpha_E = 8$ for all $E$.

%s2.3 #&#
\subsection{Recorded probing depth}
\label{recordedprobingdepth}

The translation from observed
to recorded probing depth is based on both examination
protocol and physical
characteristics of the manual probe.
The probe is scored at sequential millimeter markings
so that observed probing depths fall at or between markings.
In our study, probing depths were recorded as the greatest integer
less than or equal to the probing depth observed on the manual probe.
%For example, a recorded probing depth of 3~mm implies the observed
%depth falls between 3 and 4mm.
%The probe used in our study measures a maximum depth of 15mm so that
%a recorded
%value of 15mm corresponds to an observed value of at least 15mm.
%(
Although our protocol accommodated recorded probing depths up to 15~mm,
the largest recorded value was 8~mm.
%)

Let $U_{ijk}$ be the recorded probing depth for the
$k$th replicate
of the $j$th site for the $i$th subject. Then
%e3 #&#
\begin{eqnarray}
\label{recordeddepth} U_{ijk} &= &\cases{\lfloor T_{ijk} \rfloor, &
\quad if $0 \leq T_{ijk} < 15$,
\vspace*{3pt}
\cr
15, &\quad otherwise,}
\end{eqnarray}
where $\lfloor a \rfloor$ is the greatest integer less than or equal
to $a$.
It follows that
\begin{eqnarray*}
\pi_{u,ijk} &= &\operatorname{Pr}\bigl(U_{ijk} = u | \log(
\theta_{ij}), \bolds{\beta}_{ij}, \mathbf{X}_{ijk},
\bolds{\sigma}^2\bigr)
\\
&= &\cases{ \zeta_{u+1}, &\quad if $u = 0$,
\cr
\zeta_{u+1} -
\zeta_u, &\quad if $u = 1, \ldots, 14$,
\cr
1 - \zeta_u,
&\quad if $u = 15$,}
\end{eqnarray*}
where
%$$
$\zeta_u = \Phi
% \frac{
[\{\log(u) - \log(\theta_{ij}) -
\mathbf{X}_{ijk}^{\prime}\bolds{\beta}_{ij}\}/ \tilde{\mathbf
{X}}_{ijk}^{\prime} \bolds{\sigma}^2 ] $
%}
%{
%} \right)
%$$
and $\Phi(\cdot)$ is the standard normal distribution function.
Let
$\mathbf{V}_{ijk} = (V_{0,ijk}, V_{1,ijk}, \ldots,
V_{15,ijk})^{\prime}$
be a vector of length 16 consisting of 15 zeros and a single
one such that $V_{u,ijk} = 1$ when $U_{ijk} = u$.
Then
%$$
$\mathbf{V}_{ijk} | \log(\theta_{ij}), \bolds{\beta}_{ij}, \mathbf
{X}_{ijk}, \bolds{\sigma}^2
\sim\operatorname{Multinomial}(1; \bolds{\pi}_{ijk})$,
%$$
where $\bolds{\pi}_{ijk} = (\pi_{0,ijk}, \pi_{1,ijk}, \ldots, \pi
_{15,ijk})^{\prime}$
and the (conditional) likelihood, $L$, is given by
%e4 #&#
\begin{equation}
\label{likelihood} L \propto\prod_{i=1}^n
\prod_{j=1}^{m_i} \prod
_{k=1}^2 \prod_{u=0}^{15}
\pi_{u,ijk}^{V_{u,ijk}}.
\end{equation}

%s2.4 #&#
\subsection{Estimation and inference}\label{sec2.4}
%s2.4.1 #&#
\subsubsection{Prior specifications}\label{sec2.4.1}
For all analyses, we placed diffuse proper priors on the pocket depth
model parameters, with
$\mu\sim\mathrm{N}(0, 1000)$,
%(precision = variance$^{-1} = 0.001$),
$\sigma_b \sim\operatorname{Uniform}(0,10)$ and
$\sigma_{\varepsilon} \sim\operatorname{Uniform}(0,10)$.
Likewise, we placed $\operatorname{Uniform}(0,10)$ priors on all standard deviation parameters
of the observed probing depth model,
$\sigma_{A}$, $\sigma_{B}$, $\sigma_{C}$ and~$\sigma_{S}$.
%In a previous modeling effort, we placed vague inverse-gamma
%priors on variance parameters, but found posterior
%inference to be highly sensitive to these distributions.
%parameters
%to the so-called ``noninformative''
%inverse-gamma($\epsilon, \epsilon$) prior (e.g. $\epsilon= 0.001$)
%and demonstrates that such priors are highly informative for near-zero
%variances. \citet{Gelman2005} instead recommends
%starting with noninformative uniform prior densities on standard
%deviation
%parameters, an approach that worked well for our data.
For the DPP on the examiner effects,
we used the stick-breaking
prior of \citet{Ishwaran2001} with
$G_{E,0} = \mathrm{N}(0,1000)$,
$M_E = 6$ and $\alpha_E = 8$
for all examiners.
%We used the stick-breaking
%prior of \citet{Ishwaran2001} with $M_E$, the number of candidate
%values drawn from
%the base distribution, equal to 6 for
%all examiners. We selected $\alpha_E = 8$ based on an extensive
%sensitivity analysis
%described in Section \ref{simulationmodeldescription}.

We fit our model using WinBUGS [\citet{Lunn2000}]. We ran three
chains and assessed convergence graphically
using trace plots and modified Gelman--Rubin statistics [\citet
{Brooks1998}].
We used the batch-means method of Jones and colleagues [\citet
{Jones2006}] to assess
the precision with which posterior quantiles of agreement indices (the
endpoints of primary interest in our analysis) were estimated.
We used a burn-in of
50,500 iterations and conducted inference
based on a chain of length 10,000 from the posterior distributions of
model parameters. Additionally, we constructed point and interval
estimates of
agreement (weighted kappa, percent exact agreement and percent
agreement within 1~mm)
based on 10,000 samples from the posterior predictive distribution of
recorded probing depths.
We used an Intel Core 2 Quad CPU Windows machine for model fit with a
total run time (including monitoring of all nodes) of 1468 minutes
(approximately 24 hours) for our data.

%s2.4.2 #&#
\subsubsection{Posterior clustering inference}
\label{postclustinf}

The clustering induced by the DPP on the $\beta_{E,ij}$'s is used to
identify examiner-specific classes of biased ratings. (Henceforth the
terms `cluster' and `class' are used synonymously.)
We used the least-squares clustering approach of \citet{Dahl2006}
to identify the most likely clustering
among those sampled from its posterior distribution.
Specifically,
let $\mathbf{c}_{E,1}, \ldots, \mathbf{c}_{E,D}$ be $D$ draws from
the posterior clustering distribution of the $\beta_{E,ij}$'s. For each
clustering
$\mathbf{c}_E$ in $\mathbf{c}_{E,1}, \ldots, \mathbf{c}_{E,D}$, let
$\delta(\mathbf{c}_E)$ be an $\mathcal{L} \times\mathcal{L}$
($\mathcal{L} = \sum_{i=1}^n m_i$) association matrix with element
$\delta(\mathbf{c}_E)_{\ell\ell^{\prime}} = 1$ indicating the
examiner effects associated with sites $\ell$ and $\ell^{\prime}$
jointly classified,
and 0 otherwise.
Element-wise averaging of the collection of association\vspace*{1pt}
matrices yields the pairwise probability clustering matrix, $\Delta
_E$. Examiner E's least-squares cluster, $\mathbf{c}^{\mathrm
{LS}}_{E}$, is the
observed clustering from the Markov chain for which the squared
deviation of its association matrix, $\delta(\mathbf{c}^{\mathrm
{LS}}_{E})$, from
the pairwise probability clustering matrix, $\Delta_E$, is a minimum.
Specifically,
\begin{eqnarray*}
\mathbf{c}^{\mathrm{LS}}_{E} &= & \mathop{\arg\min}_{\mathbf{c}_E
\in\{ \mathbf{c}_{E,1}, \ldots, \mathbf{c}_{E,D} \}}\
\sum_{\ell= 1}^\mathcal{L} \sum
_{\ell^{\prime} = 1}^\mathcal{L} \bigl( \delta(\mathbf{c}_E)_{\ell\ell
^{\prime}}
- \Delta_{E,
\ell\ell^{\prime}} \bigr) ^2.
\end{eqnarray*}
The posterior clustering via Dahl's algorithm was performed on a 9-node
cluster with 72 CPUs using code written by the authors in R (version
2.8.1), and took an average of 615 minutes (approximately 10 hours) of
user time to conduct inference for a single examiner.

%s2.4.3 #&#
\subsubsection{Understanding class membership}
\label{classassociations}

For each examiner, we examined the posterior density estimates of the
$\beta_{E,ij}$'s for sites in a common class to shed light on the
magnitude and direction of bias if present. Additionally, following
\citet{Fleiss1991}, we examined the association of tooth position
(anterior versus posterior, and maxillary versus mandibular) and site
location (proximal versus \mbox{mid-tooth}, and lingual versus buccal) with
site class membership.
We compared the proportion of sites with specified tooth and location
characteristics between classes using generalized estimating equations
(to accommodate within-mouth clustering).

%s3 #&#
\section{Model evaluation}
\label{simstudy}

%s3.1 #&#
\subsection{Data simulation model}
\label{simulationmodeldescription}

Due to the extensive run time to both fit the model and conduct
posterior clustering inference for three examiners,
we conducted a simulation using a single data realization.
While generalizability of \mbox{findings} are necessarily limited, we explored
the model's ability to recover-known parameter values and measures of agreement based on draws from
the joint posterior and posterior predictive distributions, respectively.
We constructed the \mbox{simulated} data to reflect the calibration study's
experimental design and
resulting structure of the data.
Accordingly, we simulated data composed of pocket depths for each
of nine subjects using equation (\ref{pocketdepthmodel}) with $\mu=
1$, $\sigma_b = 0.2$ and
$\sigma_{\varepsilon} = 0.3$.
%However, we
%We
%specifically
%to assess point and interval estimation of model parameters and
%agreement indices based on
%the posterior predictive distribution.
For simplicity, we assumed each subject was
fully dentate, resulting in a total of 1512 simulated pocket depths.
We modeled duplicates of observed probing depth
conditional on pocket depth by specifying
examiner biases dependent on site-specific characteristics as follows:
%e5 #&#
\begin{eqnarray}
\label{simulatedobserved}
\qquad\log(T_{ijk}) &= &\log(\theta_{ij}) +
\beta_{B,ij} \cdot I(\theta_{ij} \geq4\mbox{ mm}) \cdot I(E = B)
\nonumber\\[-8pt]\\[-8pt]
&&{} + \bigl[\beta_{C_1,ij} +\beta_{C_2,ij} \cdot I(\mbox{site } j \mbox{ is DLMM})\bigr]
\cdot I(E = C) + \gamma_{ijk},\nonumber
\end{eqnarray}
where $I(\cdot)$ is a binary indicator for the stated condition (DLMM${}={}$distolingual mandibular molar),
$\beta_{B,ij} = -0.5$, $\beta_{C_1,ij} = 0.25$, and
$\beta_{C_2,ij} = -1$.
The examiner-specific effects expressed in equation (\ref{simulatedobserved}) indicate that, relative to the standard:
(1)~examiner A does not exhibit biased measuring behavior;
(2) examiner B's measurements on pockets of 4~mm or more are too shallow
by 0.5~mm on average;
and (3) overall, examiner C's measures are too deep by 0.25~mm with the
exception of
distolingual mandibular molar sites, for which measures are negatively
biased by 0.75~mm.
%Recall that the variance of $\gamma_{ijk}$ depends on the examiner (A,
%B, C or S)
%responsible for the $k$th observed probing depth for the $j$th site of
%the $i$th subject.
We further simulated observed probing depths with
$\sigma_A = 0.1, \sigma_B = 0.25, \sigma_C = 0.15$ and $\sigma_{S}
= 0.07$.
We then constructed recorded probing depths as described by equation
(\ref{recordeddepth}).
A total of 3024 (1512 pocket depths${}\times{}$2 recorded probing depths
per site)
simulated recorded probing depths comprised the final simulation data set.
%Simulated data for examiner/standard pairings AS, BS and CS were
%composed of measures from 210 periodontal sites %across 5 subjects;
%simulated data for all other examiner pairings were composed of
%measures from 126 periodontal %sites across 3 subjects.
%In all, we simulated measurements for 588 periodontal sites across 8
%subjects for examiners A, B and C, and
%756 sites across 8 subjects for the standard.
Of the sites examined by B, 82 had true depths of 4~mm or more.
Of those examined by C, 28 were from distolingual mandibular molars.

We conducted a sensitivity analysis to tune our selection of the
concentration parameter, $\alpha_E$.
Specifically, we considered values of $\alpha_E$ equal to 0.5, 1, 2,
3, 4, 5, 6, 7, 8, 9, 10 and 20, and
conducted posterior clustering inference using the method
of least-squares clustering introduced by \citet{Dahl2006} and
described in Section~\ref{postclustinf}.
For each value of $\alpha_E$ we assessed both the number of clusters
identified as well as the strength of
association between class membership and factors known to be associated
with biased measurement (e.g., deep pockets for examiner B
and distolingual mandibular molars for examiner C). We selected $\alpha
_E = 8$ based on the resulting model's ability
to recover the correct number of clusters for each examiner (1 for A
and 2 for B and C), maximum sensitivity and specificity of the
recovered cluster assignments,
and statistical significance of association of class membership with
characteristics inducing bias.

Using the simulated recorded measures as data,
we fit Model~3 as described by equations (\ref{pocketdepthmodel}),
(\ref{observedmodel}) and (\ref{likelihood}), with examiner-specific
variances and site-level examiner biases
modeled as described in Section~\ref{DPPdescription}.

%s3.2 #&#
\subsection{Simulation results}\label{sec3.2}
We summarize first our assessment of inter- and intra-examiner
agreement, as this
was the primary objective of our analysis.
For each agreement index, we report three values: the true value, derived
from true joint and marginal probabilities of recorded probing depths
based on the
model described in Section~\ref{simulationmodeldescription}; the
observed value, calculated
from the estimated joint and marginal probabilities of recorded probing
depths based on the single
simulated set of recorded probing depths derived from the model
described in Section~\ref{simulationmodeldescription};
and the median and 95\% predictive interval obtained from 10,000
estimates of each agreement measure
based on the same number of data realizations derived from the
posterior predictive distribution using
the analysis model described in Section~\ref{simulationmodeldescription}.
We begin with a detailed description of our approach to agreement
evaluation in Sections~\ref{trueagreement} through
\ref{pocketdepthagreement}.

%s3.2.1 #&#
\subsubsection{True agreement measures}\label{trueagreement}

We derived true agreement values based on theoretical
joint and marginal probabilities of recorded probing depths. Consider the
following example based on examiners B and S.
%We begin by
%constructing joint probabilities of recorded probing depth
%for examiners B and S based on the
%bivariate normal distribution of their (log transformed) observed
%measures. Specifically,
%Let $T_{B}$ and $T_{S}$ be representative (i.e. averaged
%across all subjects in the population of interest) observed probing
%depth duplicates measured by examiners
%B and S, respectively. Similarly, let $\theta$ be a representative
%pocket depth in the population.
Let $T_{B}$ and $T_{S}$ be observed probing depth duplicates measured
by examiners
B and S, respectively, for a given periodontal site with corresponding
pocket depth $\theta$.
From equation (\ref{observedmodel}), the joint distribution of $T_{B}$
and $T_{S}$ is given by
%e6 #&#
\begin{eqnarray}
\label{jointBS} %\left( \begin{array}{c}
\bigl( \log(T_{B}), % \\
\log(T_{S}) \bigr) ^{\prime} %\end{array} \right)
&\sim&\mathrm{N} (\bolds{
\mu}, \bolds{\Sigma} ),
\end{eqnarray}
where $\bolds{\mu}= (\mu- 0.5, \mu)^{\prime}$ if $\theta\geq4$~mm,
$\bolds{\mu}= (\mu, \mu)^{\prime}$ if $\theta< 4$~mm,
$\Sigma_{11} = \sigma^2_b + \sigma^2_{\varepsilon} + \sigma^2_B$,
$\Sigma_{22} = \sigma^2_b + \sigma^2_{\varepsilon} + \sigma^2_S$,
and
$\Sigma_{12} = \Sigma_{21} = \sigma^2_b + \sigma^2_{\varepsilon}$.
%$\bSigma= \left[
Defining
$\eta= \operatorname{Pr}(\theta\geq4\mbox{~mm}) =
1 - \Phi\{(\log4 - \mu)/
\sqrt{\Sigma_{12}} \}$,
%$$\eta= \operatorname{Pr}(\theta\geq4\mbox{~mm}) =
%1 - \Phi\left( \frac{\log4 - \mu}{\sqrt{\sigma^2_b + \sigma^2_{
%$1 - \eta= \operatorname{Pr}(\theta< 4\mbox{~mm})$,
the respective marginal distributions of observed probing depths are
%e7 #&#
\begin{eqnarray}
\label{marginalB} %\log(T_{B}) &\sim&\mathrm{N}(\mu- 0.5, \sigma^2_B +
\log(T_{B}) &\sim&\mathrm{N}(
\mu- 0.5, \Sigma_{11}) \eta+ %\\
%&&
\mathrm{N}(\mu, \Sigma_{11}) (1 - \eta)
\end{eqnarray}
and
%e8 #&#
\begin{eqnarray}
\label{marginalS} %\log(T_{S}) &\sim&\mathrm{N}(\mu, \sigma^2_{S} +
%+ \sigma^2_b).
\log(T_{S}) &\sim&\mathrm{N}(
\mu, \Sigma_{22}).
\end{eqnarray}
Based on the compound symmetry induced by equation (\ref{pocketdepthmodel}), distributions of observed probing depths for other
sites within the same mouth are equivalent.
Joint and marginal probabilities of recorded probing depths for other
examiner pairs are
similarly derived.

Weighted kappa, $\kappa_w$, is a chance-corrected agreement measure
that weights disagreements based on the measures' relative distance
[\citet{Cohen1968}; \citet{Fleiss1981}, pages 223--225].
Continuing with our example,
let $U_{B}$ and $U_{S}$ be recorded probing depths corresponding
to observed values $T_{B}$ and $T_{S}$.
Define $p_{u_1,u_2} = \operatorname{Pr}(U_{B} = u_1, U_{S} = u_2)$,
$p_{u_1} = \operatorname{Pr}(U_{B} = u_1)$, and $p_{u_2} = \operatorname{Pr}(U_{S} = u_2)$.
Then $\kappa_w$ is defined as
$\kappa_w = (p_{o(w)} - p_{e(w)})/(1 - p_{e(w)})$,
where
$p_{o(w)} = \sum_{u_1 = 0}^{15} \sum_{u_2 = 0}^{15} w_{u_1,u_2} p_{u_1,u_2}$,
%$p_{u_1,u_2} = \operatorname{Pr}(U_{B} = u_1, U_{S} = u_2)$,
%$p_{e(w)} = \sum_{u_1 = 0}^{15} \sum_{u_2 = 0}^{15} w_{u_1,u_2}
%p_{u_1} p_{u_2}$,
%$p_{u_1} = \operatorname{Pr}(U_{B} = u_1)$, and
%$p_{u_2} = \operatorname{Pr}(U_{S} = u_2)$.
%p_{o(w)} &= &\sum_{u_1 = 0}^{15} \sum_{u_2 = 0}^{15} w_{u_1,u_2}
and
%p_{e(w)} &= &\sum_{u_1 = 0}^{15} \sum_{u_2 = 0}^{15} w_{u_1,u_2}
$p_{e(w)} = \sum_{u_1 = 0}^{15} \sum_{u_2 = 0}^{15} w_{u_1,u_2}
p_{u_1} p_{u_2}$.
We use the common weighting scheme
$w_{u_1,u_2} = 1 - \{(u_1 - u_2)^2/(N-1)^2 \}$,
%$$w_{u_1,u_2} = 1 - \frac{(u_1 - u_2)^2}{(N-1)^2},$$
where $N$ is the total number of categories (16 in this case).

We also constructed measures of percent exact agreement, $P_{\mathrm{exact}}$,
and percent agreement within 1~mm, $P_{\pm1}$, where
%In our example,
%$P_{\mathrm{exact}}$
%and
%$P_{\pm1}$ are defined as
%P_{\mathrm{exact}} &= &\sum_{u = 0}^{15} \operatorname{Pr}(U_{B} =
%U_{S} = u)
$P_{\mathrm{exact}} = \sum_{u = 0}^{15} \operatorname{Pr}(U_{B} =
U_{S} = u)$
and
\begin{eqnarray*}
P_{\pm1} &= &P_{\mathrm{exact}} + \sum_{u=1}^{15}
\bigl\{ \operatorname{Pr}(U_{B} = u, U_{S} = u-1) + \mbox
{Pr}(U_{B} = u-1, U_{S} = u) \bigr\}.
\end{eqnarray*}
%
%$P_{\pm1} = P_{\mathrm{exact}} +
%The probabilities used to construct true values of $\kappa_w$, $P_{
%are derived based on
%the relationship between observed and recorded probing depth described
%in Equation (\ref{recordeddepth}), and
%from Distributions
%(\ref{jointBS}), (\ref{marginalB}) and (\ref{marginalS}).

We constructed the true value of $\kappa_w$ for each examiner pair
using the true values of $p_{u_1}$, $p_{u_2}$ and $p_{u_1,u_2}$
obtained from the joint and marginal distributions shown in (\ref
{jointBS})--(\ref{marginalS}) (or analagous distributions for other
examiners), and the relationship between observed and recorded probing
depths described by equation (\ref{recordeddepth}).
In a~similar manner, we constructed true values of $P_{\mathrm{exact}}$ and $P_{\pm1}$.
The resulting agreement values for each examiner pair are reported in
the column labeled ``Truth''
in Table~\ref{simulationagreement}.
%
%t1 #&#
%
\begin{table}[b]
\tabcolsep=2.75pt
\caption{Simulation agreement results. A, B, C and S are the three
examiners and standard. PD${}={}$pocket depth.
AS, BS and CS results are based on 210 site-level measures from 5
subjects. Results for all other
pairings are based on 126 measures from 3 subjects.
Observed results are point estimates obtained from the simulated data
set, and results obtained from the
posterior predictive distribution are medians and 95\% predictive intervals}\label{simulationagreement}
\begin{tabular*}{\tablewidth}{@{\extracolsep{\fill}}@{}lccd{3.1}cc d{3.1}cc ccc@{}}
\hline
&\multicolumn{3}{c}{\textbf{Truth}} &\multicolumn{3}{c}{\textbf{Observed}} &\multicolumn{3}{c}{\textbf{Post pred}}\\[-6pt]
&\multicolumn{3}{c}{\hrulefill} &\multicolumn{3}{c}{\hrulefill} &\multicolumn{3}{c}{\hrulefill}
\\
& & \multicolumn{1}{c}{\textbf{\%}} & & & \multicolumn{1}{c}{\textbf{\%}} & & & \multicolumn{1}{c}{\textbf{\%}} &
\\
\textbf{Pair} & \multicolumn{1}{c}{$\bolds{\kappa_w}$} & \multicolumn{1}{c}{\textbf{agree}} & \multicolumn{1}{c}{\textbf{\% $\bolds{\pm1}$}}
              & \multicolumn{1}{c}{$\bolds{\kappa_w}$} & \multicolumn{1}{c}{\textbf{agree}} & \multicolumn{1}{c}{\textbf{\% $\bolds{\pm 1}$}}
              & \multicolumn{1}{c}{$\bolds{\kappa_w}$} & \multicolumn{1}{c}{\textbf{agree}} & \multicolumn{1}{c}{\textbf{\% $\bolds{\pm1}$}}
\\
\hline
AS &0.890 &72.2 &99.5 &0.902 &66.7 &100.0 &0.846 &66.2 &99.1 \\
&&&&&& &(0.765, 0.927) &(55.7, 76.7) &(95.7, 100.0)\\
BS &0.693 &44.9 &89.3 &0.454 &51.0 &83.3 &0.669 &45.2 &90.0 \\
&&&&&& &(0.550, 0.819) &(34.8, 57.1) &(80.5, 96.2)\\
CS &0.664 &31.4 &81.3 &0.591 &31.0 &79.0 &0.613 &31.4 &79.1 \\
&&&&&& &(0.498, 0.770) &(22.4, 43.3) &(66.7, 90.0)\\
AB &0.683 &44.0 &88.5 &0.429 &48.4 &88.9 &0.633 &43.7 &88.9 \\
&&&&&& &(0.474, 0.801) &(31.8, 57.1) &(77.0, 96.8)\\
AC &0.659 &31.7 &80.8 &0.709 &28.6 &77.8 &0.586 &31.0 &78.6 \\
&&&&&& &(0.449, 0.762) &(19.8, 45.2) &(62.7, 91.3)\\
BC &0.547 &26.5 &70.5 &0.454 &34.1 &81.7 &0.497 &26.2 &70.6 \\
&&&&&& &(0.332, 0.694) &(16.7, 38.9) &(54.8, 85.7)\\
AA &0.872 &68.1 &99.0 &0.825 &64.3 &98.4 &0.835 &65.1 &98.4 \\
&&&&&& &(0.730, 0.923) &(50.8, 77.8) &(93.7, 100.0)\\
BB &0.559 &35.8 &79.8 &0.344 &33.3 &83.3 &0.619 &42.1 &85.7 \\
&&&&&& &(0.441, 0.789) &(29.4, 55.6) &(73.0, 95.2)\\
CC &0.719 &43.1 &84.5 &0.845 &54.0 &88.9 &0.819 &48.4 &91.3 \\
&&&&&& &(0.712, 0.907) &(35.7, 63.5) &(81.0, 98.4)\\
SS &0.911 &77.2 &99.8 &0.884 &70.6 &100.0 &0.872 &72.2 &100.0 \\
&&&&&& &(0.776, 0.947) &(59.5, 84.1) &(96.8, 100.0)\\
A/PD &0.910 &77.0 &99.8 &&& &0.873 &71.8 &99.3 \\
&&&&&& &(0.815, 0.939) &(63.6, 79.6) &(97.6, 100.0)\\
B/PD &0.703 &45.9 &90.1 &&& &0.696 &46.8 &91.2 \\
&&&&&& &(0.611, 0.830) &(38.6, 56.8) &(84.0, 95.9)\\
C/PD &0.669 &30.9 &81.9 &&& &0.629 &31.3 &80.1 \\
&&&&&& &(0.545, 0.779) &(23.5, 42.0) &(69.6, 89.5)\\
S/PD &0.936 &83.8 &100.0 &&& &0.917 &80.3 &100.0 \\
&&&&&& &(0.875, 0.963) &(72.9, 86.5) &(99.5, 100.0)\\
\hline
\end{tabular*}
\end{table}

%s3.2.2 #&#
\subsubsection{Observed agreement}\label{observedagreement}
Additionally, for each examiner pair we constructed empirical estimates
of $\kappa_w$, $P_{\mathrm{exact}}$ and $P_{\pm1}$ based
on estimated joint and
marginal probabilities of recorded probing depths from the simulated
data set
described in Section~\ref{simulationmodeldescription}.
These values are reported in the column labeled ``Observed'' in Table~\ref{simulationagreement},
and are obtained by
using sample proportions to estimate the probabilities required to
construct the
agreement measures.

%s3.2.3 #&#
\subsubsection{Posterior predictive agreement estimates}
\label{postpredagreement}

We also obtained point and interval estimates of agreement for each
examiner pair from 10,000 data realizations obtained from the posterior
predictive distribution
based on the Bayesian analysis model described in Section~\ref{simulationmodeldescription}.
Specifically, for each data set simulated from the posterior predictive
distribution, we estimated
joint and marginal probabilities of recorded probing depths based on
sample proportions,
and subsequently constructed estimates of $\kappa_w$, $P_{\mathrm{exact}}$ and $P_{\pm1}$ for each examiner pair.
These values are reported in the column labeled ``Post pred'' in Table~\ref{simulationagreement}.

%s3.2.4 #&#
\subsubsection{Examiner agreement with pocket depth}
\label{pocketdepthagreement}
We define examiner agreement with true pocket depth
as the value of the agreement measure achieved when the collection
of recorded probing depths associated with a given examiner,
$\{U_{ijk}\}$, are compared to the corresponding values of true pocket
depths censored according to the rule described in equation (\ref{recordeddepth}).
We derived both true measures
of agreement based on theoretical joint and marginal probabilities as
well as point and interval estimates
of agreement resulting from the 10,000 data realizations from the
posterior predictive distribution
and summarize these results in
Table~\ref{simulationagreement}.
Because truth is not observable, we omit
a measure of ``Observed'' agreement with true pocket depth.

%s3.2.5 #&#
\subsubsection{Simulation agreement summary}\label{sec3.2.5}
Although results based on a single simulated data set preclude generalizability,
the observations summarized herein are meant to provide a ``first
look'' at model performance.
Agreement measures for the simulated
data are summarized in Table~\ref{simulationagreement}.
%Note the `observed' and `posterior predictive' estimates are
%equivalently constructed from
%joint and marginal probabilities estimated from appropriate sample
%proportions.
We observe that estimates based on the posterior predictive
distribution recover agreement indices' true values with
all 95\% predictive intervals containing the truth, although we can
make no claims pertaining to coverage.
%However, with results based on a single simulated data set, we can
%make no claims pertaining to coverage or generalizability.
%Estimation based on the posterior predictive distribution accurately
%recovers agreement indices' true values with
%all 95\% predictive intervals containing the truth.
Still, there are advantages in using repeated draws from the posterior
predictive
distribution to construct these indices.
The ability to obtain interval estimates correctly accounting for
the correlation of pocket depths in the same mouth and between
duplicate readings is a strength.
Furthermore, only the model-based approach provides an estimate of
agreement between each examiner and true pocket depth.
%, arguably the measure of greatest interest.
Finally, since the posterior predictive draws are sampled from a
distribution derived from
a model utilizing the complete data, the pooling of information yields
improved power to make
inferential statements about individual examiners.

%Finally, since the posterior predictive draws are sampled from a
%distribution derived from
%a model utilizing the complete data,
%the subsequently constructed agreement estimates (most notably $
%tend to be smoothed in the direction of the truth.

%The latter observation is consistent with conclusions made by
%Guggenmoos-Holzmann and Vonk (\citeyear{Guggenmoos1998}),
%who show that Cohen's kappa underestimates chance-corrected agreement
%(but can occasionally be too large) when
%the distributions of chance and systematic ratings differ, or when
%systematic disagreement is involved.
%To mitigate these distorting effects, they suggest using more
%informative study designs incorporating simultaneous %assessment of
%intra- and inter-examiner variation.
%We observe this biased estimation in our simulation. For example, true
%$\kappa_w$ for examiner B and the standard was %0.693, while its
%observed value based on the simulated data was 0.454 (Table
%primary source of this discrepancy, we simulated 1000 data
%realizations using the model %described in Section
%and constructed 1000 $\kappa_w$ estimates based on the resulting data
%subsets specific to examiner B with the standard. %The average $
%0.623 respectively. However, %estimated $\kappa_w$ based on the
%posterior predictive distribution from Model 3 is 0.669 with endpoints
%of the 95\% %predictive interval equal to 0.550 and 0.819,
%demonstrating our model's corrective strength.

%s3.2.6 #&#
\subsubsection{Model parameters}\label{sec3.2.6}
Table~\ref{simulationparams} shows posterior estimates of all model
parameters (except the $\beta_{E,ij}$'s).
%;Model 3 recovers the true parameter values with high accuracy.
Supplementary Figure~1 [\citet{HISL13}] shows posterior density
estimates of the $\beta_{E,ij}$'s for examiners A, B and C. The posteriors for A effects are strongly unimodal and
centered at zero, indicating no bias.
In contrast, posterior densities for examiner B effects indicate two
modes, one at 0 and another at $-$0.5.
Similarly, posterior densities for examiner C effects identify two
modes, one at 0.25 and the second at $-$0.75.
The locations of these modes are consistent with the data simulation
model described by equation (\ref{simulatedobserved}).

%t2 #&#
%
\begin{table}
\tablewidth=230pt
\tabcolsep=4pt
\caption{Simulation model posterior parameter estimates (median and
95\% predictive interval).
For each examiner, we report the true and estimated number of classes
as determined by the method of
least-squares clustering}\label{simulationparams}
\begin{tabular*}{\tablewidth}{@{\extracolsep{\fill}}ld{1.2}d{1.12}@{}}
\hline
\textbf{Parameter} & \multicolumn{1}{c}{\textbf{Truth}} & \multicolumn{1}{c@{}}{\textbf{Posterior estimate}}\\
\hline
$\mu$ &1 &1.03\ (0.80, 1.18) \\
$\sigma_b$ &0.2 &0.19\ (0.11, 0.40) \\
$\sigma_{\varepsilon}$ &0.3 &0.29\ (0.28, 0.30) \\
$\sigma_A$ &0.1 &0.11\ (0.09, 0.13) \\
$\sigma_B$ &0.25 &0.24\ (0.22, 0.28) \\
$\sigma_C$ &0.15 &0.15\ (0.12, 0.17) \\
$\sigma_{S}$ &0.07 &0.08\ (0.07, 0.10) \\
A classes &1 &\multicolumn{1}{c@{}}{1\tabnoteref{tt1}\hphantom{0\ (0.07, 0.10)}}\\
B classes &2 &2 \\
C classes &2 &2 \\
\hline
\end{tabular*}
\tabnotetext{tt1}{Two additional classes included a single site each.}
\end{table}

A number of the $\beta_{E,ij}$'s posterior density estimates are
bimodal, suggesting, perhaps, that
examiner effects sometimes suffer from nonidentifiability.
This is most pronounced for examiner pairings without the standard,
situations in which
there is less information to ``anchor'' true pocket depth ($\theta_{ij}$).
%While the
%availability of examiner/standard pairings allows for correct
%identification of the modes, the
%bimodality of certain posterior density estimates reflects the
%uncertainty in correcting for the
%discrepancy between the observed probing depth and the currently
%sampled pocket depth in the Markov chain.
Still, the modes and their relative heights inform on rating behavior:
examiner A's measures are unbiased;
examiner B's measures are most often unbiased but
sometimes negatively biased; and examiner C's measures are most often
positively biased but
sometimes negatively biased.
This information together with the least-squares cluster assignment of
the examiner effects
provides a picture of both the
magnitude of bias as well as those factors influencing bias.

In our simulation, the least-squares clustering for examiner A effects
identified a single dominant class
consistent with our simulation of no bias for examiner A measures.
Two additional classes were identified for examiner A, each comprising
a single site.
On closer inspection, these correspond to the only cases in which A
records a probing depth of 0.
In these situations, $\log(T_{ijk})$ will take on large negative
values when the Markov chain for $T_{ijk}$ samples small positive values.
The Markov chain for the corresponding examiner effect will likewise
sample large negative values and the posterior density
estimate is subsequently diffuse and negatively skewed with a single
mode at 0.
Because these posterior distributions are so markedly different from
the norm,
%(see Figure \ref{simbetaplots}),
the examiner effects for these sites are assigned to singleton classes.

The least-squares clusterings for examiners B and C both identified two
classes---a single dominant class corresponding to the highest
mode of the posterior density estimates
%shown in Figure \ref{simbetaplots},
and a second class corresponding to the
subordinate mode. Specifically, examiner B's classes comprised 539 and
49 site-specific examiner effects.
%We further investigated the significance of the association between
%class membership and factors known to influence examiner bias using
%generalized estimating
%equations.
Recalling examiner B's biased measures for deep pockets,
10\% (54 of 539)
of sites associated with the larger class
were deep versus 57\% (28 of 49) in the smaller class ($p < 0.0001$),
although the corresponding sensitivity was weak [Sens${}={}$Prob(subordinate class assignment${}|{}$deep site)${}\doteq 28/82 = 34$\%].
For examiner C, the classes comprised 547 and 41 site-specific examiner effects.
One-half percent (3 of 547) of sites associated with the larger class
were distolingual mandibular molars
versus 61\% (25 of 41) in the smaller class ($p < 0.0001$), and the
sensitivity was excellent
[Sens${}={}$Prob(subordinate class assignment${}|{}$DLMM)${}\doteq 25/28 = 89$\%].

%
%
%
%
%
%
%
%
%

%
%
%
%

%s4 #&#
\section{Application to calibration training data}
\label{application}

We fit the reference model, Model 0, and Models 1, 2 and 3 (described
in Sections~\ref{model1} through \ref{DPPdescription})\break
to the calibration training data. We compared model fit using
$\mathrm{DIC}_3 =\break -4 \mathrm{E}_{\vartheta} [ \log f(\mathbf
{U}|\vartheta) | \mathbf{U} ] + 2 \log\hat{f}(\mathbf{U})$,
as described by \citet{Celeux2006}, where
%$DIC$_3$ as described by \citet{Celeux2006}, where
$\vartheta$ is a vector of model parameters,
\[
\hat{f}(\mathbf{U}) = \prod_{i=1}^{n}
\prod_{j=1}^{m_i} \prod
_{k=1}^{2} \hat{f}(U_{ijk})
\]
and
%$$\hat{f}(U_{ijk}) \doteq\mbox{E}_{\vartheta} \left[ f(U_{ijk}|
$\hat{f}(U_{ijk})$ approximates $\mathrm{E}_{\vartheta} [
f(U_{ijk}|\vartheta)|\mathbf{U} ]$,
the predictive density averaged over the MCMC run.
The fit was dramatically improved for Model 3 (DIC$_3$ for Models~0
through 3 were 4560.11,
4402.13, 4129.07 and 3381.83).

%Model &DIC3\\ \hline
%0 &4560.11\\
%1 &4402.13\\
%2 &4129.07\\
%3 &3381.83\\ \hline

%t3 #&#
%
\begin{table}
\tabcolsep=0pt
\caption{Examiner calibration training agreement results.
A, B, C and S are the three examiners and standard. PD${}={}$pocket depth.
Italicized estimates and 95\% confidence intervals (CIs) are
obtained from the observed data as described in Section \protect\ref{observedagreement}.
Nonitalicized estimates (medians and 95\% predictive intervals) are
obtained as described in Sections \protect\ref{postpredagreement} and \protect\ref{pocketdepthagreement}. The number of subjects and sites examined by
examiners $E$ and $E^{\prime}$ is
given by $n_{EE^{\prime}}$ and $\mathcal{L}_{EE^{\prime}}$, respectively}\label{SUNYAgreementResults}
\begin{tabular*}{\tablewidth}{@{\extracolsep{\fill}}lcd{3.0}ccc@{}}
\hline
\textbf{Pair} & \multicolumn{1}{c}{$\bolds{n_{EE^{\prime}}}$} & \multicolumn{1}{c}{$\bolds{\mathcal{L}_{EE^{\prime}}}$}
              & \multicolumn{1}{c}{$\bolds{\kappa_w}$}        & \multicolumn{1}{c}{\textbf{\% agree}}
              & \multicolumn{1}{c}{\textbf{\% $\bolds{\pm1}$}}
\\
\hline
AS &5 &180 &$\mathit{0.713}$ &$\mathit{62.2}$ $\mathit{(36.1,
88.4)}$ &$\mathit{94.4}$ $\mathit{(83.3, 100.0)}$\tabnoteref{t31}\\
& & &0.793 (0.687, 0.893) &58.9 (47.2, 69.4) &95.0 (89.4, 98.9)\\
BS &5 &156 &$\mathit{0.666}$ &$\mathit{48.7}$ $\mathit{(24.6, 72.8)}$ &$\mathit{87.8}$ $\mathit{(70.6, 100.0)}$\tabnoteref{t31}\\
& & &0.641 (0.485, 0.796) &43.6 (32.1, 55.1) &88.5 (78.2, 94.9)\\
CS &5 &180 &$\mathit{0.691}$ &$\mathit{42.8}$ $\mathit{(34.2, 51.4)}$ &$\mathit{92.2}$ $\mathit{(82.4, 100.0)}$\tabnoteref{t31}\\
& & &0.709 (0.586, 0.836) &47.2 (35.6, 58.9) &92.2 (83.3, 97.2)\\
AB &3 &108 &$\mathit{0.629}$ &$\mathit{45.4}$ $\mathit{(28.0, 62.7)}$ &$\mathit{81.5}$ $\mathit{(49.6, 100.0)}$\tabnoteref{t31}\\
& & &0.601 (0.420, 0.772) &40.7 (28.7, 54.6) &85.2 (73.2, 94.4)\\
AC &3 &96 &$\mathit{0.585}$ &$\mathit{43.8}$ $\mathit{(0.0, 88.5)}$\tabnoteref{t32}
&$\mathit{87.5}$ $\mathit{(63.7, 100.0)}$\tabnoteref{t31}\\
& & &0.622 (0.443, 0.793) &43.8 (30.2, 58.3) &87.5 (76.0, 95.8)\\
BC &3 &120 &$\mathit{0.615}$ &$\mathit{46.7}$ $\mathit{(13.2, 80.2)}$ &$\mathit{80.8}$ $\mathit{(59.0, 100.0)}$\tabnoteref{t31}\\
& & &0.602 (0.433, 0.768) &45.0 (32.5, 57.5) &88.3 (76.7, 95.8)\\
AA &2 &60 &$\mathit{0.896}$ &$\mathit{73.3}$\tabnoteref{t33} &$\mathit{98.3}$ $\mathit{(77.2, 100.0)}$\tabnoteref{t31}\\
& & &0.839 (0.685, 0.930) &61.7 (43.3, 78.3) &98.3 (88.3, 100.0)\\
BB &2 &72 &$\mathit{0.581}$ &$\mathit{55.6}$ $\mathit{(43.8, 67.3)}$ &$\mathit{94.4}$ $\mathit{(35.6, 100.0)}$\tabnoteref{t31}\\
& & &0.644 (0.409, 0.829) &45.8 (30.6, 63.9) &88.9 (73.6, 98.6)\\
CC &2 &78 &$\mathit{0.728}$ &$\mathit{79.5}$ $\mathit{(59.4, 99.5)}$ &$\mathit{97.4}$ $\mathit{(67.4, 100.0)}$\tabnoteref{t31}\\
& & &0.792 (0.616, 0.904) &61.5 (43.6, 76.9) &97.4 (88.5, 100.0)\\
SS &1 &30 &$\mathit{0.971}$ &$\mathit{80.0}$\tabnoteref{t34} &$\mathit{100.0}$\tabnoteref{t34}\\
& & &0.866 (0.664, 0.966) &73.3 (46.7, 93.3) &100.0 (93.3, 100.0)\\
A/PD &8 &444 &0.811 (0.734, 0.896) &61.9 (52.5, 70.5) &95.5 (91.2,
98.4)\\
B/PD &8 &456 &0.689 (0.593, 0.811) &45.6 (36.0, 55.7) &90.8 (82.9,
95.6)\\
C/PD &8 &474 &0.738 (0.650, 0.844) &48.3 (38.2, 58.7) &94.1 (87.3,
97.7)\\
S/PD &7 &546 &0.931 (0.869, 0.974) &81.3 (69.2, 91.2) &100.0 (98.9,
100.0)\\
\hline
\end{tabular*}
\tabnotetext{t31}{Upper bound truncated at 100.}
\tabnotetext{t32}{Lower bound truncated at 0.}
\tabnotetext{t33}{95\% CI not estimable because cluster-specific point estimates were equal.}
\tabnotetext{t34}{95\% CI not estimable with a single cluster.}
\end{table}

Table~\ref{SUNYAgreementResults} shows agreement results for the
calibration training data.
Italicized values are those reported by \citet{Hill2006} and are
constructed as described
in Section~\ref{observedagreement}. Nonitalicized values are medians and
95\% predictive intervals constructed from 10,000
draws from the posterior predictive distribution as described in Sections~\ref{postpredagreement} and \ref{pocketdepthagreement}.
Evaluation of the precision with which quantiles of agreement indices
were estimated from the MCMC yielded posterior standard errors no
larger than 0.008 [\citet{Jones2006}].
%Results based on the posterior predictive distribution indicate
%dramatic efficiency gains
%relative to estimation based on data subsets and asymptotics.
We observe reductions in the widths of nearly all agreement interval
estimates, likely due to the pooling of information across examiners
and subjects in a single model.
%For percent agreement, the average
%Results based on the posterior predictive distribution indicate
%half-width of 95\% confidence intervals was 23.2\% while %the average
%half-width of 95\% predictive intervals
%was 14.8\%, equivalent to a 36\% reduction in width. For percent
%agreement within 1~mm, average half-width
%improved from 17.3\% for confidence intervals to 7.8\% for predictive
%intervals, equivalent to a 55\% reduction.
%These improvement estimates are conservative since numerous confidence
%intervals based on the observed data were %truncated at 0 or 100\%.
Furthermore, interval estimates were available for all agreement
indices despite the small number of
subjects (clusters) examined by examiner pairs, a major limitation for
interval estimation based on
traditional asymptotics.

\begin{figure}

\includegraphics{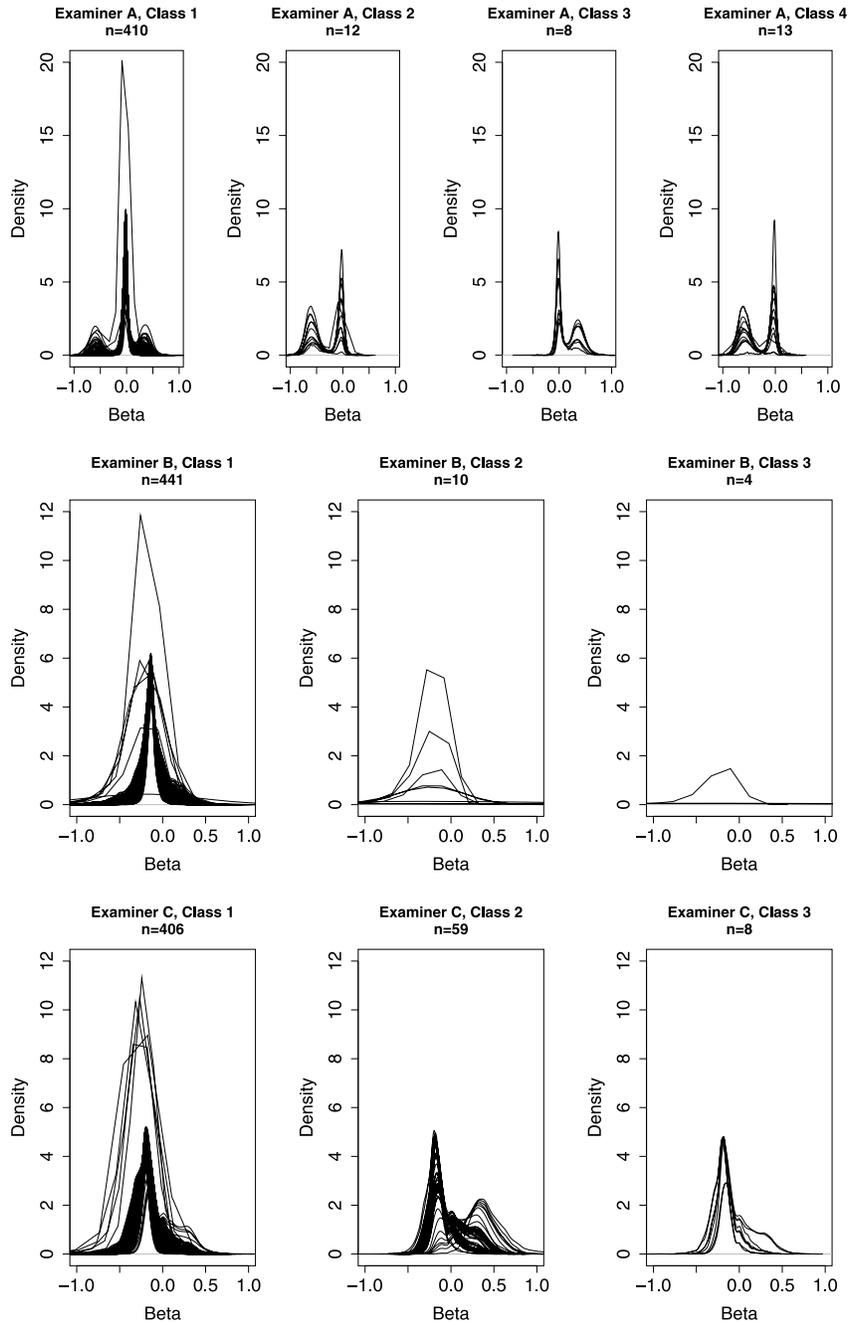}

\caption{Examiner-specific posterior density estimates of bias
parameters ($\beta$'s) for least-squares clusters based on
examiner calibration training data.}\label{sunybuffbetas}
\end{figure}

Figure~\ref{sunybuffbetas} shows posterior density estimates of the
$\beta_{E,ij}$'s for each examiner across classes
identified by the least-squares clustering algorithm.
An additional class comprising a single site was identified for each examiner.
(The posterior density estimates of the corresponding examiner effects
for these sites are not shown in
Figure~\ref{sunybuffbetas}.) These singleton classes were all cases in
which the examiner recorded
a probing depth of 0. The associated bias parameters' posterior density
estimates are diffuse and negatively skewed,
a behavior we observed in our simulation for similar data.

%f1 #&#
%

Based on similarities among posterior density estimates, we collapsed
into a single group those sites in
classes: 2 and 4 for examiner A; 2 and 3 for examiner B; and 1 and 3
for examiner C.
This resulted in posterior clustering inference based on three classes
for examiner A, two for examiner B, and
two for examiner C (excluding singleton classes).
Examiner A's measures are predominantly unbiased (class 1), but with
some evidence of both negative (class 2)
and positive (class 3) bias. Examiner B's measures are overall mildly
negatively biased (class 1), but
14 sites in class 2 are cases in which examiner B's recorded probing
depth is 0.
In contrast, only 12 of the 441 sites in class 1 are associated with a
recording of 0 by examiner B.
Examiner C's measures are overall mildly negatively biased (class 1),
but a number of sites are measured with positive bias (class 2).

%f2 #&#
%
\begin{figure}

\includegraphics{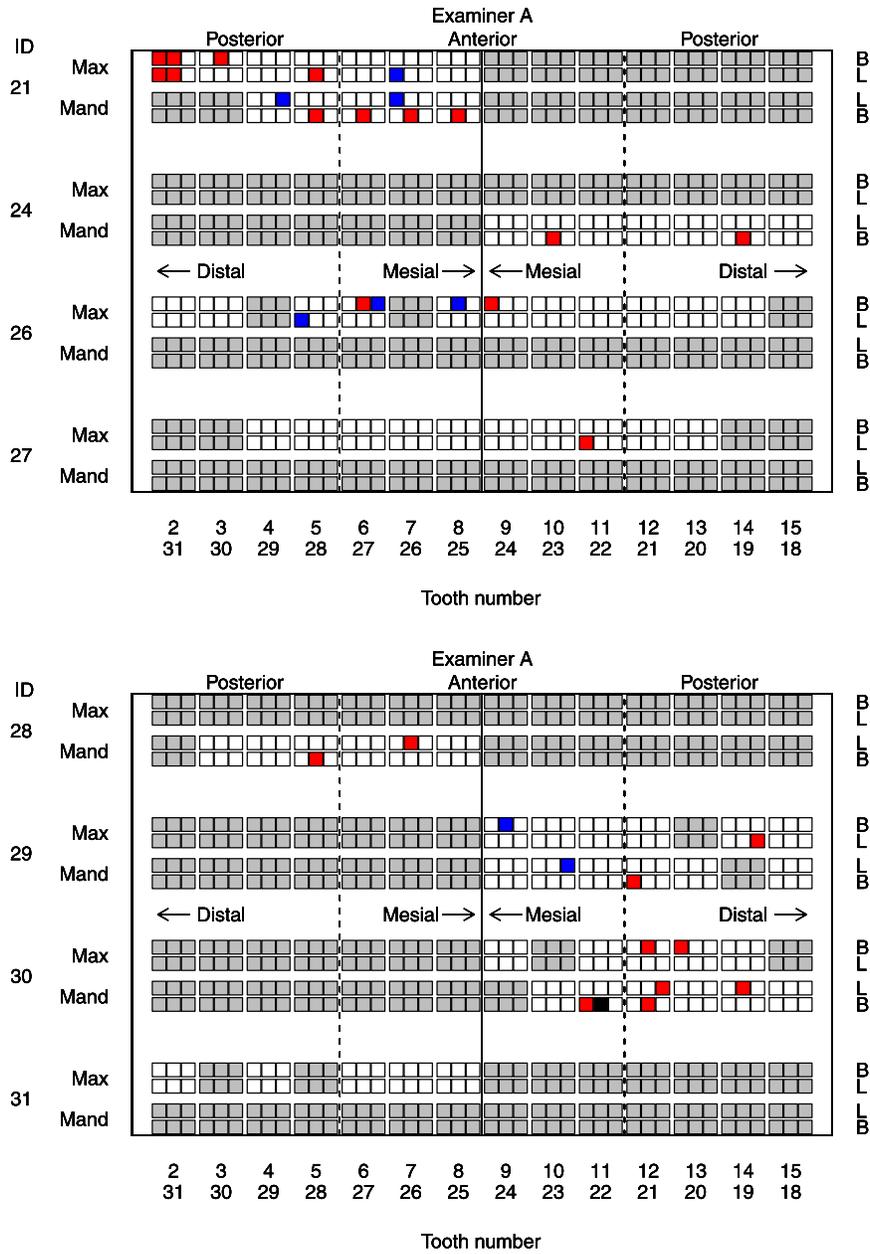}

\caption{Examiner calibration training exercise examiner A posterior
class structure. White${}={}$class~1; Red${}={}$class~2; Blue${}={}$class~3;
Black${}={}$Singleton class; Gray${}={}$not observed by examiner. Anterior versus
posterior teeth are indicated by the vertical dashed line. For each
tooth, the mesial and distal sites are those closest to and furthest
from the midline of the dental arch, indicated by the solid vertical
line. Abbreviations: Max${}={}$Maxillary; Mand${}={}$Mandibular; B${}={}$Buccal; L${}={}$Lingual.}\label{coolplotA}
\end{figure}

%Following \citet{Fleiss1991}, we assessed the association of
%tooth position (anterior versus posterior, and
%maxillary versus mandibular) and site location (proximal versus
%mid-tooth, and lingual versus buccal) with class %membership.
Figure~\ref{coolplotA} shows the distribution of class membership
across the mouth for
sites examined by examiner A [\citet{Slate2012}]. Using the
approach described in Section~\ref{classassociations}, for examiner A
we observed a significantly larger proportion of: mid-tooth sites in
class 2 versus class 1 (64\% versus 31\%, $p = 0.030$); buccal sites in
class 2 versus class 1 (68\% versus 49\%, $p = 0.016$); and sites
associated with anterior teeth in class 3 versus class 1 (75\% versus
49\%, $p = 0.028$).
Recalling for examiner A that class 1 reflects no bias, class 2
reflects negative bias, and class 3 positive bias, we conclude that
examiner A is significantly more likely to be negatively biased for
mid-tooth and buccal sites and more likely to be positively biased for
anterior teeth.
Based on similar analyses for examiner B comparing class 2 to class 1,
we observe a greater proportion of mid-tooth sites (100\% versus 31\%)
and sites located on mandibular teeth (93\% versus 70\%, $p = 0.052$).
Again, recalling for examiner B that class 1 sites are negatively
biased, and class 2 sites have recorded depths of 0, we conclude
examiner B displays an overall negative bias in measuring behavior
relative to the standard, and is more likely to measure a depth of 0~mm
for mid-tooth and mandibular sites. For examiner C, there is marginal
evidence of a larger proportion of anterior teeth in class 2 versus
class 1 (49\% versus 24\%, $p = 0.052$). Recall examiner C's class 1
and class 2 sites are positively and negatively biased, respectively,
relative to the standard. We conclude examiner C's measures are overall
negatively biased, but tend to be positively biased on anterior teeth.

We also examined the relationship between
class membership and pocket depth. Specifically, we calculated the
median of the
posterior distribution of $\theta_{ij}$ and assessed the significance
of its association
with class membership.
Examiner B is significantly more likely to
record a probing depth of zero for more shallow sites ($p = 0.012$),
and examiner C is significantly positively biased for deeper sites ($p
= 0.0007$).

\section{Discussion}\label{discussion}

In this manuscript we develop a novel approach to inter- and
intra-examiner agreement
using a semi-parametric Bayesian model with a Dirichlet process prior on
model parameters capturing examiner biases, accommodating
%We use a hierarchical modeling approach that accommodates the
dependence among measures obtained from the same unit, as well as the
dependence between duplicate measures made on the same experimental subunit.
At the suggestion of a referee, we fit an alternative model for pocket
depth [equation (\ref{pocketdepthmodel})] with an additional
tooth-level random effect. We observed modest improvement in fit
relative to Model 3 with a reduction in $\mathrm{DIC}_3$ ($\Delta\mathrm{DIC}_3 =
13.29$), but there was no meaningful change in posterior inference
(results not shown).
Recently, a number of authors have demonstrated spatial correlation
among measures obtained from periodontal sites within the same mouth,
with higher correlation among measures obtained from sites
closer together than from those further apart [\citet
{Reich2007}]; we speculate the tooth-level random effect captures some
of this spatial heterogeneity. An interesting extension of our approach
would investigate improvements in model fit and subsequent inference by
specifically modeling spatial correlation among sites in equation (\ref{pocketdepthmodel}).

Our analysis has several implications with respect to the design of
experiments intended to measure examiner agreement.
First, the discovery of sample items where examiners demonstrate
greater difficulty with agreement suggests over-sampling within these
discovered classes in follow-up calibration studies. \mbox{Furthermore,} it
may be possible to reduce the sample sizes needed to determine
agreement within specified precision bounds because the model borrows
strength across examiner pairs.
%It would be illuminating to determine, in the context of our model,
%the relative contribution of data from the various %examiner pairs to
%the precision of agreement indices.
Finally, we observed in our simulated data set that agreement indices
(in particular, $\kappa_w$) tend to be smoothed in the direction of
the truth.
Although our simulation does not permit generalization, this
observation is consistent with conclusions made by Guggenmoos-Holzmann
and Vonk [\citet{Guggenmoos1998}] who
show that Cohen's kappa may be biased when
examiners disagree systematically.
To mitigate this bias, they suggest using more informative study
designs incorporating simultaneous assessment of intra- and
inter-examiner variation, a characteristic of the design used in our
examiner calibration exercise.

%The latter observation is consistent with conclusions made by
%Guggenmoos-Holzmann and Vonk (\citeyear{Guggenmoos1998}),
%who show that Cohen's kappa underestimates chance-corrected agreement
%(but can occasionally be too large) when
%the distributions of chance and systematic ratings differ, or when
%systematic disagreement is involved.
%To mitigate these distorting effects, they suggest using more
%informative study designs incorporating simultaneous %assessment of
%intra- and inter-examiner variation.
%We observe this biased estimation in our simulation. For example, true
%$\kappa_w$ for examiner B and the standard was %0.693, while its
%observed value based on the simulated data was 0.454 (Table
%primary source of this discrepancy, we simulated 1000 data
%realizations using the model %described in Section
%and constructed 1000 $\kappa_w$ estimates based on the resulting data
%subsets specific to examiner B with the standard. %The average $
%0.623 respectively. However, %estimated $\kappa_w$ based on the
%posterior predictive distribution from Model 3 is 0.669 with endpoints
%of the 95\% %predictive interval equal to 0.550 and 0.819,
%demonstrating our model's corrective strength.

%Although we apply our method to the analysis of periodontal probing
%depth measures,
Our approach is not limited to periodontal data applications. For
example, a~common measure of anti-tumor activity in cancer clinical
trials is tumor response, measured on an ordinal scale but derived from
a continuous measure of the percentage of tumor shrinkage from baseline
in (potentially) multiple target lesions in the same subject
[\citet{Eisenhauer2009}]. This endpoint is typically measured by
expert reading of CT or MRI scans by trained radiologists. One can
envision a calibration exercise in which radiologists are trained to
measure response, but scan assessments may be biased based on (for
example) tumor location or scan quality. When pooling across examiner
pairs is appropriate, our hierarchical model provides refined inference
for calibration data that yields greater precision and identification
of classes of units measured with similar bias, contributions that
enhance the knowledge gained and enable subsequent targeted examiner training.

%A by-product of the Dirichlet process prior on the examiner bias
%parameters is the posterior clustering of
%effects. Using Dahl's method of least-squares clustering
%classes of sites that are measured with similar bias. Furthermore, by
%examining the relationship of site- and %tooth-level characteristics
%with
%class membership, we identified factors significantly associated with
%bias for each examiner.
%Thus our model provides important information that can be used to
%target examiner training
%thereby potentially improving data quality in periodontal research.

\section*{Acknowledgments}
The authors thank Dr. Carlos Salinas and the clinical core members of the
South Carolina Center of Biomedical Research Excellence for Oral Health
at the Medical University of South Carolina
%SC Oral Health COBRE
for use of the periodontal data.
Additionally, the authors thank two reviewers and an Associate Editor
for helpful comments that substantially improved
the manuscript.

\begin{supplement}%[id=suppA]
\stitle{A semi-parametric Bayesian model of inter- and intra-examiner
agreement for periodontal probing depth: Supplementary Figure\\}
\slink[doi]{10.1214/13-AOAS688SUPP} %[doi,text={...}] - jei reikia
%suskaldyti doi
\sdatatype{.pdf}
\sfilename{aoas688\_supp.pdf}
\sdescription{Posterior density estimates of bias parameters ($\beta_{E, ij}$'s) for
examiners A, B and C based on the simulation model described in Section~\ref{simulationmodeldescription}.}
\end{supplement}

% imsref loaded by linak, 2013-12-12 09:51:01
%

\printaddresses

\end{document}